\documentclass[conference]{IEEEtran}
\IEEEoverridecommandlockouts
\usepackage{cite}
\usepackage{amsmath,amssymb,amsfonts}
\usepackage{algorithmic}
\usepackage{graphicx}
\usepackage{textcomp}
\usepackage{xcolor}
\usepackage{verbatim}
\def\BibTeX{{\rm B\kern-.05em{\sc i\kern-.025em b}\kern-.08em
    T\kern-.1667em\lower.7ex\hbox{E}\kern-.125emX}}

\begin{document}

\title{Precise Indoor Positioning Based on UWB and Deep Learning\\
}


\author{Chenyu Wang, Zihuai Lin\\
	School of Electrical and Information Engineering, The University of 
	Sydney, Australia\\
	Emails:	zihuai.lin@sydney.edu.au. 
}

\maketitle

\begin{abstract}
We examined UWB-based indoor location in conjunction with a fingerprint technique in this work. We built a connection between the measured and real distances for the UWB indoor positioning system. This connection is used to produce a distance database that may be used to generate fringerprints. We created a BP neural network to classify the target node to the relevant fringerpint using the distance database. Our suggested deep learning technology considerably enhances location accuracy when compared to existing trilateration systems.
\end{abstract}
\smallskip

\section{Introduction}\label{sec:introduction}

Nowadays, the concept of mobile networks is embedded in daily lives, which provides a positive environment for developing applications based on positioning and localization technologies. 

Outdoor location and tracking technology has matured into a mature application that is widely recognized and applied all over the world. The Global Location System (GPS) and the BeiDou Navigation Satellite System (BDS) are two examples of outdoor positioning and tracking applications. These systems can provide mobile users with high-accuracy location services in outdoor environments.
However, GPS has a number of fundamental flaws that make it unsuitable for indoor location applications. For example, there is a high chance that GPS signals are unable to travel through reinforced concrete structures \cite{WooSunkyu2011AoWi,LeeYang-Won2008AGst}. As a result, when a user is within a building, tunnel, or subterranean construction, GPS cannot offer accurate position information.

To overcome systemic limitations of GPS and provide satisfactory localization services to users with indoor conditions, Ultra-wideband (UWB) is introduced by researchers. UWB is sort of signal that has a frequency band between 3.1 to 10.6 GHz, which is suitable for real-time indoor positioning applications. Additionally, there are multiple methods that can generate UWB signal, such as frequency-hopping, time-hopping, and direct-sequence \cite{WinMoeZ2009HaAo}, which increases the flexibility of applying UWB to indoor positioning.

To further increase the accuracy of the indoor positioning via UWB, Machine Learning (ML) algorithms, such as decision trees, random forests, KNN and soft voting algorithms are investigated in \cite{Anqi} to determine the locations, among which the soft voting algorithm performs the best. In this work, we develop deep learning based indoor positioning algorithms to 
improve system performance in terms of accuracy of positioning. 

The remainder of the paper is organized as follows. Section \ref{sec:litreivew} introduces the background about indoor positioning and the related techniques. Section \ref{sec:III}  introduces the system model and positioning methods used in the paper. Section \ref{sec:IV} describes the experiment settings, data collections and positioning process of the experiment. Section \ref{sec:V} analyzes the experimental results. The last Section gives a summary of this work and the future work plan.

\section{Background\label{sec:litreivew}}

\subsection{Indoor positioning methods}
The most common approaches for indoor positioning are Angle of Arrival (AOA), Received Signal Strength Indicator (RSSI) \cite{position2,position1}, and Time of Arrival (TOA) \cite{NessaAhasanun2020ASoM,leng2020implementation}. 


The Angle of Arrival approach is based on the angle measurements. According to \cite{BergenMarkH2018Ttio}, there are two main coordinate frames for AOA ranging. 
The first frame is the so-called local frame which is a spheroidal coordinate frame with origin of local anchor node. This type of coordinate frame uses the anchor node to measure and determine the direction of the target and the distance between the target and itself. After measurement, the obtained data in spheroidal coordinate frame is processed and converted to the second type of coordinate frame called global frame. Global frame consists of a standard three-dimensional coordinate frame with $x$, $y$, $z$ axis. Therefore, the current location of the target can be detected. Theoretically, two anchor nodes with known locations are enough for an AOA positioning system.

The RSSI method measures the path loss of the signal transmitted from the target to an anchor node. By measuring the received signal strength and comparing with transmitted signal strength, the value of strength difference can be calculated. Then, this value becomes the input of a theoretical propagation attenuation model which converts measured value to useful distance parameters. These parameters are further used to determine the location of the target \cite{HuiLiu2007SoWI}. 

Time of Arrival is another approach which can be used to obtain the distance between the target and anchor node. According to \cite{DardariDavide2009RWUB}, there are two kinds of TOA techniques. The first type is called one-way TOA, which requires prefect synchronization of system clocks at both the target and anchor node. In one-way TOA, the target sends a clock information to the anchor node. Therefore, the anchor node can compare the time difference between transmitting and receiving, which can be used to calculate the propagation distance from the target to anchor node. The second type is called two-way TOA ranging, which does not need a synchronization of system clocks. The anchor node sends a request message to the target and record the sending time locally. The target then transmits an acknowledgement back to the anchor node after receiving the request message. By doing this way, the anchor node can also calculate the time difference and associate distance parameter in the end.

Compared with AOA, RSSI and TOA approaches require at least three anchor nodes deployed at fixed locations \cite{NessaAhasanun2020ASoM}. The Trilateration algorithm is a commonly used method for calculating the target node location. 
For a two dimension environment, three anchor nodes are needed to determine the location of a target node. 
Each anchor node 
measures the distance between the target and the anchor node itself. Three ranging circles can be produced based on the received data by sharing measurement data with each other. The target's putative position is defined by the intersection of these circles. However, positioning accuracy can be degraded due to several practical issues \cite{NessaAhasanun2020ASoM}. For example, environmental noises might interfere with the ranging signal, lowering the precision of measurement during the ranging process. Aside from that, range accuracy might be impacted by the Non-Line of Sight (NLOS) problem, which occurs when an impediment blocks the direct sight from the anchor node to the target. Besides the above mentioned methods, emerging radio sensing techniques, such as \cite{GI1,GI2,GI3,GI4} can also be used for indoor positioning.

\subsection{Fingerprint in Indoor Positioning}

In fingerprint positioning, the whole positioning area is divided into several equal parts \cite{NessaAhasanun2020ASoM}. There are two major phases in fingerprint methods: offline phase and online phase. In the offline phase, the system records location information of all reference points distributed in a testing zone, which sets up a local database to store the ‘signature’ of the entire testing zone \cite{YanJun2018HKBM}. In the online phase, the ranging techniques such as AOA, RSSI, or TOA can be applied to obtain real-time data of the target node. By matching the signature and estimation, the best location information can be determined \cite{YanJun2018HKBM}.

\subsection{Machine Learning Algorithms}

The machine learning algorithm \cite{ML_AmBC,ML_Peng} is one of the most popular and common methods utilized nowadays to implement fingerprint theory.	There are many distinct strategies used in machine learning algorithms; two well-known techniques are Decision Tree (DT) and Support Vector Machine (SVM). Decision tree can gather fingerprint with different operations in offline and online phases. According to \cite{ChanamaLummanee2018Acod}, the location parameters should be measured and recorded before setting up a decision tree model. The entire testing zone can be divided into multiple grids. 
The characteristics of location data in each grid can be analysed to train the decision tree algorithm so that different decision labels can be created. After learning, the characteristic of new location data of target is compared with decision labels to classify the current data into the data cluster with the most similar label. Furthermore, the best pair of coordinates can be determined after input data going through the decision tree and completing iterative classification process \cite{YimJaegeol2008Iadt}.

Support Vector Machine is another proven technique which can applied for indoor positioning application combining with the fingerprint concept. Compared to other ML algorithms, SVM is one of the supervised algorithms which maintains a relatively low level of computation complexity \cite{AdegeAbebeBelay2018AIaO}. It aims to train the algorithm based on existing data, and classify new input based on calculating difference between the training model and new input. These basic features make SVM suitable for solving regression problem. However, specific training logic of SVM leads to drawbacks such as time consuming and large memory occupation \cite{NessaAhasanun2020ASoM}.

\subsection{Deep Learning and Neural Networks}

Deep learning and neural networks, in addition to traditional machine learning methods, can be used for indoor positioning applications. The Non Line-of-Sight (NLOS) issue \cite{NLOS1,NLOS2,NLOS3,NLOS4,NLOS5} is one of the most serious shortcomings of indoor positioning. When an obstacle blocks the direct link between a target and an anchor node, signals sent by the target may still be received by the anchor node. However, range parameters such as RSSI and TOA might be less precise, if not erroneous, reducing the overall system's positioning accuracy \cite{AlyaaThamirSalim2019RBBF}.

Deep learning and neural networks are essentially extensions of machine learning. This type of algorithms may be treated as a network made up of many nerve cells. When an input is provided to the algorithm, one or more nerve cells will capture it in order to create an associated output depending on internal rules. Upper cells' outputs will be used as inputs by the cells below. This procedure will continue until the final nerve cell produces an output that is deemed the system's ultimate results.


Each nerve cell is given a weight to assure the algorithm's learning capabilities. Before moving on to the next cell, the input data needs to be adjusted depending on the relevant weights. Algorithms can then learn the characteristics of data and modify weights to optimize the result. Such functions can be accomplished in a variety of ways. For instance, according to \cite{AlyaaThamirSalim2019RBBF}, Radial Basis Function (RBF) is integrated with neural networks. To begin, get the weights from the neural network's input and hidden layers. Then, using the hidden layer output and weight values determine the final result. There are numerous primary aspects that might impact the positioning system's learning efficiency and accuracy.
As stated in \cite{AlyaaThamirSalim2019RBBF}, the required training time is a crucial factor. Longer learning time with enough amount of training samples can improve system performance significantly. Additionally, the number of neurons in hidden layer is another crucial factor which can significantly impact the accuracy \cite{BrahimElBoudani2020IDLT}. However, the larger the number of neurons, the higher the memory capacity required. As a result, there is a trade-off to be addressed between system load and placement precision.

The Back Propagation (BP) Neural Network is one of the most appealing methods employed and developed in terms of neural network implementation in indoor positioning applications. The BP neural network, like a normal feed-forward neural network structure with a supervised learning algorithm, has at least three layers divided into three categories.
%
%
The input layer and output layer are first two categories of layers included in the BP neural network structure, where the number of nerve cells for both input and output layers can be adjusted depending on the requirements of specific applications. Apart from that, the BP neural network could extend the size of hidden layer with relatively high level of flexibility. The special structure and relatively high flexibility of the BP neural network provide significant advantages. According to \cite{ZhuWenqi2021Roip}, the BP neural network has a strong ability on non-linear representation and prediction. Additionally, benefiting from strong self-learning ability, a BP neural network can theoretically learn and simulate all the abstract functional relationship between multiple variables \cite{WangWei2021Roip}.

\section{System Model and Methods\label{sec:III}}



\begin{figure}[htb]
    \centering
    \includegraphics[width=.4\textwidth]{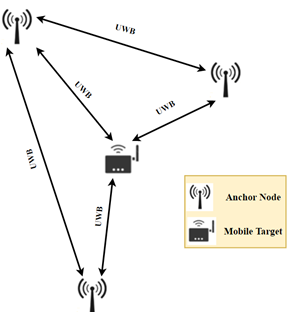}
    \caption{Precise Indoor Positioning System Model}
    \label{fig.my_label3_1}
\end{figure}

The investigated system consists of a mobile target and three anchor nodes as shown in Fig.\ref{fig.my_label3_1}. 
The communication method utilized by all four devices is UWB-based wireless communication techniques. The TOA ranging approach is used to measure the distance between each individual anchor node and the mobile target. All the measured data is sent to one anchor node which is connected to a computer for data processing. 

\subsection{Solutions and Methods}


In this work, we use mathematics tools and real measured distance values to generate a database of the location data at different coordinates of the testing zone.

Based on our observation, there is a linear relationship between the measured distance and the real distance between an anchor node and a mobile target. This relationship can be described by eq. \ref{linearEQ}, where $d_{real}$ and $d_{test}$ represent the real and test distance between an anchor node and the target, respectively, and $a$ and $b$ are two real values.
\begin{equation}\label{linearEQ}
     d_{test} = a * d_{real} + b.
\end{equation}
To determine the values of $a$ and $b$, at least two reference points with known coordinates are required. The $d_{real}$ is known for two reference points based on coordinates, and the distance $d_{test}$ from each reference point to an anchor node can be measured during the experiment. By applying data into eq. \ref{linearEQ}, we can calculate the values of $a$ and $b$. In our work, to minimize the error between the measured distance and real distance, we do the experiment several hundreds times to get the values of $a$ and $b$. Based on these values and eq. \ref{linearEQ}, we can be then generate training database for our developed BP neural network.

\subsubsection{Measurement Calibration (MC)}

Measurement Calibration is a tentative method utilised in this project in order to eliminate the negative impact caused by self-defect of the UWB devices. The main purpose of this method is to explore and analyse the potential relationship between measured distance values and real values. There are several reference points selected from the testing zone with known position coordinates. According to the known coordinates of the target and anchor nodes, the precise distance between each individual anchor node and the mobile target can be calculated by 
\begin{equation}
    d_{real} = \sqrt{(x_A-x_T)^2 +(y_A-y_T)^2},
\end{equation}
where ($x_A$, $y_A$) and ($x_T$, $y_T$) are the position coordinates of an anchor node and the mobile target, respectively. We then do experiment to measure the distance between this anchor node and the mobile target. For one target position, we measure many times to get averaged values of $a$ and $b$ for eq. \ref{linearEQ}. This way provides the possibility of compensating measurement error caused by self-defect of equipment numerically.

\subsubsection{Two-point Diagonal Training (TDT)}

\begin{figure}[htb]
    \centering
    \includegraphics[width=.5\textwidth]{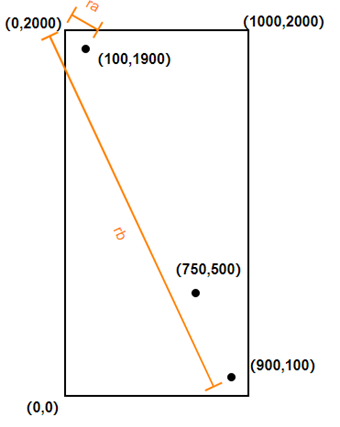}
    \caption{Schematic Diagram of TDT Method}
    \label{fig.my_label3_4_2}
\end{figure}
In the TDT method, the reference points are selected along the diagonal line of the testing zone as shown in Fig. \ref{fig.my_label3_4_2}. In the figure, ‘$ra$’ and ‘$rb$’ represent the shortest distance measured from reference points to the anchor node located at the vertex of the rectangular testing zone with coordinates (0, 2000). In order to explore the influence of reference point positions on neural network training set, there are three reference points tested. As the case that only two reference points are required each time to calculate the values of $a$ and $b$, three reference points are divided into two separate groups shown in Table \ref{tab.my_label3_1}. 
\begin{table}[h]
\centering
\caption{TDT Groups}
\label{tab.my_label3_1}
\begin{tabular}{|c|c|}
    \hline
    Group 1 & Group 2 \\
    \hline
    (750, 500); (900, 100) & (100,1900); (900,100) \\
    \hline
\end{tabular}
\end{table}

\subsubsection{Two-point Non-diagonal Training (TNT)}

\begin{figure}[htb]
    \centering
    \includegraphics[width=.5\textwidth]{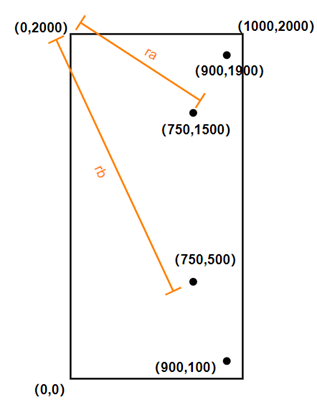}
    \caption{Schematic Diagram of TNT Method}
    \label{fig.my_label3_4_3}
\end{figure}
Two-point Non-diagonal Training is another developed rule of selecting reference points based on distance simulation approach to generate the distance database. As three anchor nodes are placed at (0, 0), (0, 2000) and (1000,2000), the points located at the upper left region of the testing zone has better measurement accuracy, because points in this region has higher probability that be within 1000 mm to all three anchor nodes. However, points in the bottom right region has a relatively high probability to be over 1000 mm to two anchor nodes. Therefore, choosing more reference points located in the bottom right region could help the neural network to learn comprehensive features of the bottom right region, which makes it possible to improve the performance. 
In the TNT rule, the two reference points are selected non-diagonally in the testing zone as shown in Fig. \ref{fig.my_label3_4_3}. Similar to the TDT rule mentioned earlier, ‘$ra$’ and ‘$rb$’ also represent the shortest distance measured from reference points to the anchor node located at vertex of the rectangular testing zone with coordinates (0, 2000). In addition, four reference points are also selected and split into two groups as described in Table \ref{tab.my_label3_2}.
\begin{table}[h]
\centering
\caption{TNT Groups}
\label{tab.my_label3_2}
\begin{tabular}{|c|c|}
    \hline
    Group 1 & Group 2 \\
    \hline
    (900, 1900); (900, 100) & (750,1500); (750,500) \\
    \hline
\end{tabular}
\end{table}

\subsubsection{Coalescent Training (CT)}

\begin{figure}[htb]
    \centering
    \includegraphics[width=.5\textwidth]{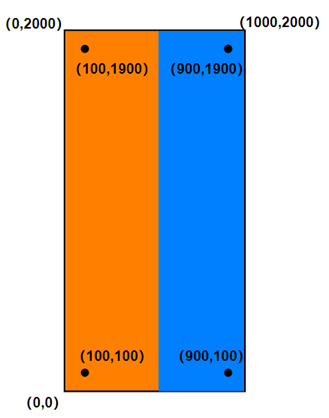}
    \caption{Schematic Diagram of CT Method}
    \label{fig.my_label3_4_4}
\end{figure}
Coalescent Training is a training method developed based on both the TNT and TDT rules. In this rule, the entire testing zone is divided into two subzones. For the left-hand side of the testing subzone, the Two-point Diagonal Training method is applied. The values of $a$ and $b$ are calculated based on four reference points as shown in Fig. \ref{fig.my_label3_4_4}. The first set of $a$ and $b$ values is calculated based on the measured data of reference points (100, 1900) and (900, 100) to the anchor node at (0, 2000). Another set of $a$ and $b$ values is calculated based on the measured data of reference points (100, 100) and (900, 1900) to the anchor node at (0, 0). Finally, these two sets of $a$ and $b$ values are averaged to become the final parameters applied to generate the distance database for the left-hand side of the testing subzone. On the other hand, the TNT rule is applied to the blue region as shown in Fig. \ref{fig.my_label3_4_4}. The data of reference points (900, 100) and (900, 1900) is selected to calculate the set of $a$ and $b$ values with respect to the anchor node at (1000, 2000). By implementing the distance database generation process to both sides of the testing zone with corresponding set of $a$ and $b$ values, the training distance database can be generated and integrated before passing to the BP neural network.

\section{Experiment\label{sec:IV}}

\subsection{Data collection}

\begin{figure}[htb]
    \centering
    \includegraphics[width=.5\textwidth]{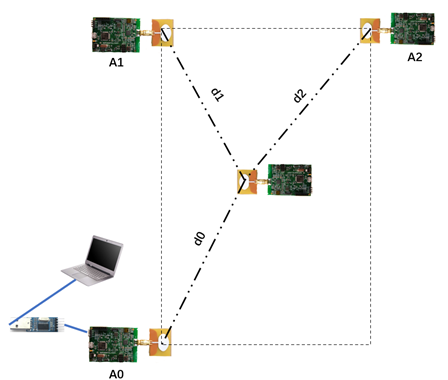}
    \caption{The UWB-based indoor positioning system}
    \label{fig.my_label4_2}
\end{figure}

In this project, we used a UWB hardware system to determine the location of the target node. The coordinate for each location of the target node is computed via geometric calculation based on collected TOA distance data. The system is illustrated in Fig. \ref{fig.my_label4_2}.

The system consists of four UWB devices, three of which are anchors nodes, i.e., A0, A1, and A2 with known locations. The testing zone is of size 2x1 meters. Another UWB device is used as a mobile target node. 
%
Based on the TOA technique, the distances from the target node to all the three anchor nodes,i.e, d0, d1, d2 can be measured. 



After data collection, the trilateration method is used to determine the location coordinate of the target node based on the distance measurements. The calculated coordinate is then compared with the known  coordinate. The distance between these two coordinates can be calculated and utilized as a measurement of system performance. 


\subsection{Distance database generation flow chart}

\begin{figure}[htb]
    \centering
    \includegraphics[width=0.5\textwidth]{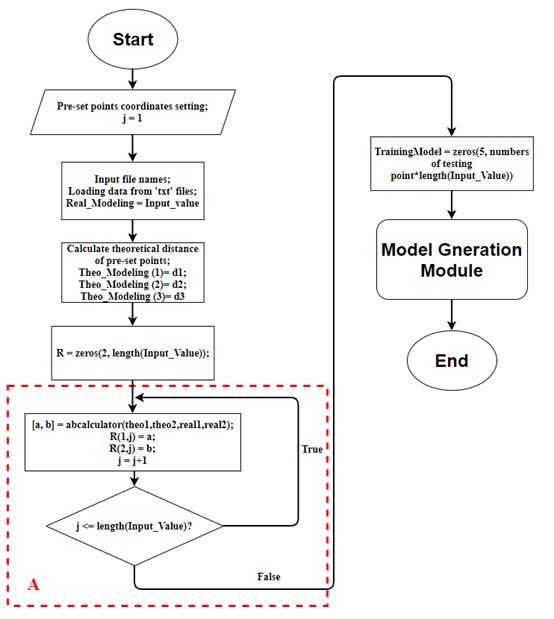}
    \caption{Flow Chart of Generating Training Model}
    \label{fig.my_label4_3a}
\end{figure}

A flow chart of the training model in the distance database generation process is described in Fig.  \ref{fig.my_label4_3a}.
As shown in the flow chart, the coordinates for pre-set points are declared in the beginning of the process. Based on the distance measurements from the target node at each pre-set point to all anchor nodes, the values of $a$ and $b$ of eq. \ref{linearEQ} can be determined.
These values are then used for generating distance database for the entire testing zone.


\subsection{Positioning process}
\begin{figure}[htb]
    \centering
    \includegraphics[width=0.5\textwidth]{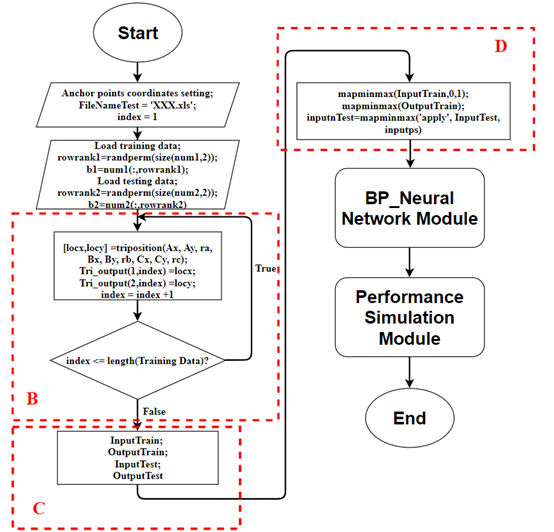}
    \caption{Flow Chart of System Simulation}
    \label{fig.my_label_4_3b}
\end{figure}

 Fig. \ref{fig.my_label_4_3b} illustrates the flow chart for the entire positioning process.
The overall positioning process can be divided into five steps. In the first step, the coordinates of three anchor points are declared. Apart from that, the data measurements are loaded to program memory before starting to execute following steps. In the second step, the trilateration method is used to calculate the location of the target node. (This is the performance of positioning system without using the neural network). In the third step, the loaded data for both training and testing are assigned to corresponding input and output sets before feeding to the neural network module. In the fourth step, all training and testing data needs to be normalized so that all data can be unified in the same level of magnitude. 
The last step is to build the BP neural network, training and testing the location of the target node based on the BP neural network. 


\section{Results and Analysis\label{sec:V}}

\subsection{Results of Measurement Calibration}

\begin{figure}[htb]
    \centering
    \includegraphics[width=.5\textwidth]{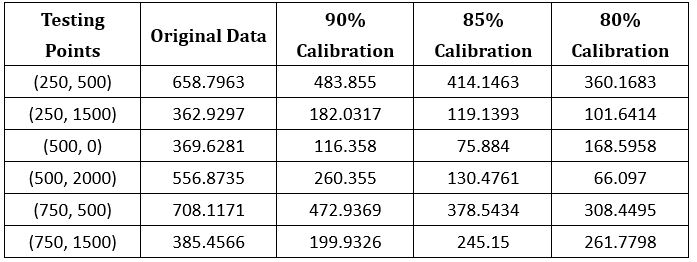}
    \caption{Mean Distance Error Distance using the Trilateration method}
    \label{fig.my_label5_1A}
\end{figure}

\begin{figure}[htb]
    \centering
    \includegraphics[width=.5\textwidth]{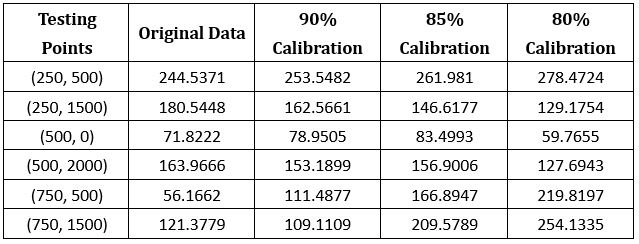}
    \caption{Mean Distance Error using the BP Neural Network}
    \label{fig.my_label5_1b}
\end{figure}

There are six points selected and tested during the experiments. We observed that when real distance is longer than 1000 mm, the measured distance is always greater than the actual real distance. Therefore, measurement calibration is introduced in order to compensate the errors. We scale down the measured data to $K$ percent of its original value if the original value is greater than 1000 mm, where $K$ is set to 90, 85, and 80 respectively. The calibrated data is then passed to the Trilateration module as well as the BP neural network module for calculating the target node's coordinate. We then calculate the distance error between each reference point and the calculated target point. The results for the Trilateration method and the BP Neural Network method are shown in Figs. \ref{fig.my_label5_1A} and \ref{fig.my_label5_1b}, repectively.

From Fig.\ref{fig.my_label5_1A} we can see, measurement calibration can improve the error performance for the system using conventional trilateration methods. With the enhancement of calibration, the mean error distance of most of testing points decrease, which means that the accuracy of the positioning system is improved. However, for the system using the BP Neural Network method, the error performance does not improve as shown in Fig. \ref{fig.my_label5_1b}. One probable explanation is that the suggested calibration procedure is so radical that some of the measured data's properties are over-corrected and  the basic learning algorithm with the BP neural network is highly referring on the typical characteristics of data. On the other hand, by comparing the results in both Figs. \ref{fig.my_label5_1A} and \ref{fig.my_label5_1b}, it is clear that the positioning system using the BP neural network provides better performance than the system based on conventional trilateration methods. 

Although measurement calibration method tested in the project generally improves error performance of the positioning system, this method is not reliable in a general situation due to its limitations. First, this calibration method can lead to over correction issues, which can be verified by the results of testing point (250, 500), (750, 500), and (750, 1500) in Fig. \ref{fig.my_label5_1b}. With the enhancement of calibration, the mean error distance increased, which means the BP neural network is misled by over corrected training data. second, in a general situation, measured error can be caused due to various of factors. The effect of self-defect of equipment can only be one of the major factors. The interference of operation environment, the different complexity of testing field can also impact the final accuracy of positioning system. Therefore, this measurement calibration method can be hardly applied to a general case in practice.


\subsubsection{Results of TDT}

\begin{figure}[htb]
    \centering
    \includegraphics[width=.5\textwidth]{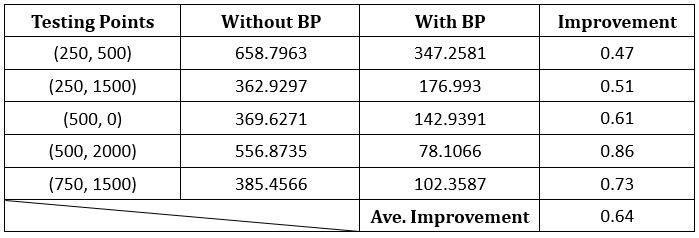}
    \caption{TDT Mean Error Distance of Group 1}
    \label{fig.my_label5_2a}
\end{figure}

\begin{figure}[htb]
    \centering
    \includegraphics[width=.5\textwidth]{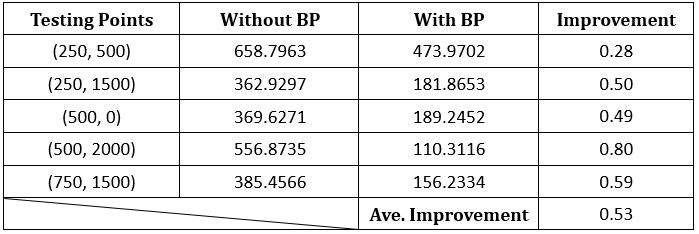}
    \caption{TDT Mean Error Distance of Group 2}
    \label{fig.my_label5_2b}
\end{figure}

As described earlier, the TDT method is proposed to generate the distance database. We selected two groups of reference points for testing. 
The first group consists of points (750, 500) and (900, 100), while the second group consists of points (100, 1900) and (900, 100). To minimise the errors, each reference point is measured 300 times. The error distance is calculated and averaged in the end of the experiment. The results of mean error distance for both groups are listed in Figs. \ref{fig.my_label5_2a} and \ref{fig.my_label5_2b}.


It is clear that for both groups, the BP neural network approach brings a significant improvement compared with the conventional Trilatriation method as shown in the improvement column of Figs. \ref{fig.my_label5_2a} and \ref{fig.my_label5_2b}. However, by comparing the results of both groups, it can be observed that with deep learning, the 2nd group gives worse performance than the first group. This maybe due to the self-defect of the equipment, the measurement accuracy is relatively low when the target node is very close to the anchor node. This is the case that one of the reference points in group 2 is close to the anchor node, the accuracy of measured distance cannot keep within an acceptable range. These damaged data cannot provide the neural network module with the right features of the testing zone. Therefore, the performance descends when data of group 2 is applied.

This experiment indicates that the reference points should be selected closing to the central region of the testing zone instead of the anchor nodes. 

\subsubsection{Results of TNT}
\begin{figure}[htb]
    \centering
    \includegraphics[width=.5\textwidth]{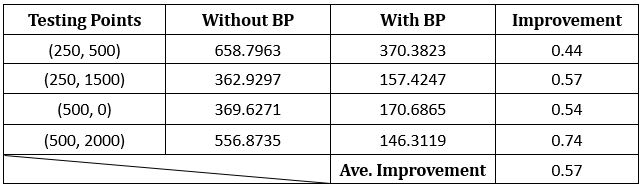}
    \caption{TNT Mean Error Distance of Group 1}
    \label{fig.my_label5_3a}
\end{figure}

\begin{figure}[htb]
    \centering
    \includegraphics[width=.5\textwidth]{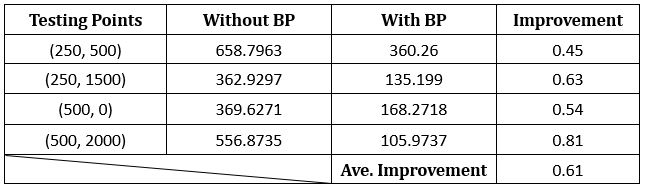}
    \caption{TNT Mean Error Distance of Group 2}
    \label{fig.my_label5_3b}
\end{figure}

In this part, we use the Two-point Non-diagonal Training approach to select reference points for training the neural network. Similar to the TDT case, we also select two groups of reference points for testing. The reference points selected in the first group are (900, 1900) and (900, 100), while reference points selected in the second group are (750, 500) and (750, 1500). After collecting the measured data of all four reference points, we use the similar process to obtain the error distance performance. The results are shown in Figs. \ref{fig.my_label5_3a} and \ref{fig.my_label5_3b}, respectively. 
It can be seen that reference points in group 2 give better performance.

\subsection{Results of Coalescent Training (CT)}
\begin{figure}[htb]
    \centering
    \includegraphics[width=0.5\textwidth]{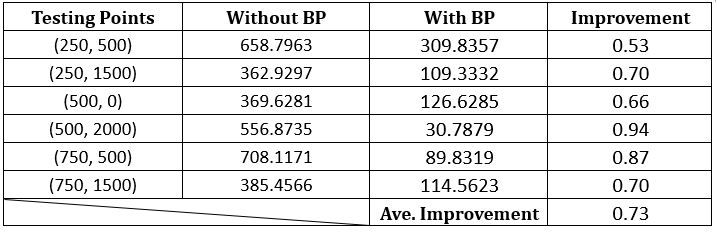}
    \caption{CT Mean Error Distance}
    \label{fig.my_label5_4a}
\end{figure}

As described in Section \ref{sec:III}, the coalescent training is an approach which integrates the features of both the TDT and TNT methods. By applying this approach to the BP neural network, we obtain the following results shown in Fig. \ref{fig.my_label5_4a} 
\begin{figure}[htb]
    \centering
    \includegraphics[width=0.5\textwidth]{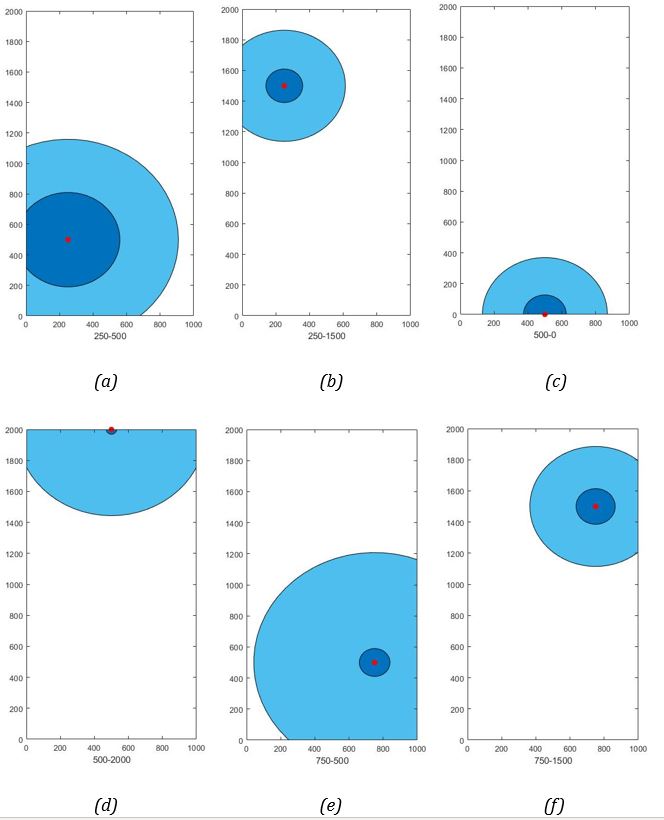}
    \caption{CT performance}
    \label{fig.my_label5_4b}
\end{figure}

To observe the performance of the system intuitively, there are six plots generated and shown in Fig. \ref{fig.my_label5_4b} based on the obtained results. The precise position coordinate of each individual testing point is represented by the red dot in the Figure.  The dark blue region for each testing point represents the mean error range using the BP neural network algorithm for positioning. On the other hand, the bright blue region represents the mean error range using the trilateration method. It can be seen from Figs. \ref{fig.my_label5_4a} and \ref{fig.my_label5_4b} that the CT-based BP neural network brings a significant improvement on positioning accuracy. Compared to the conventional trilateration method, the CT-based BP neural network approach brings about 73 percent improvement on system accuracy in average.

\section{Conclusion\label{cha:conclusion}}

In this paper, we investigated UWB-based indoor positioning. For the UWB indoor positioning system, we created a link between the measured and true distances. This relationship is used to create a distance database for usage as a fringerprint. Based on the distance database, we developed a BP neural network to categorize the target node to the relevant fringerpint. When compared to traditional trilateration approaches, our proposed deep learning methodology significantly improves location accuracy.

In terms of future development, the neural network's potential will be studied and evaluated, so that the indoor positioning system based on UWB and neural network may be more accurate. To begin with, the fingerprint database employed in the present system model is built using characteristics with a modest number of points. The more data utilized, according to the features of fingerprint and neural network algorithms, the higher the performance can be attained. As a result, it is worthwhile to employ more points to create the fingerprint database. Aside from that, further ways of picking reference points with common characteristics should be investigated in the future. This will increase the neural network's performance even more. Finally, in the future, additional types of neural network algorithms, such as convolutional neural network and radial basis function algorithms, can be simulated and evaluated. Furthermore, deep learning algorithms and denoising techniques developed in our existing works, e.g., \cite{IREALCARE1,IREALCARE2,IREALCARE3,IREALCARE4,IREALCARE5} can be used to improve the accuracy. Privacy concerns for location-based services \cite{privacy1,privacy2,privacy3,privacy4} and cellular networks \cite{NC,JNCC, UAV_THz, cellular1,cellular2,cellular3,MIMO_capacity,UAVdownlink,OC-isit} and sensor networks \cite{network_capacity,DistributedRateless,DistributedRateless2,DistributedRateless3,NC4,NC5,WRN,WRN2,WRN3, Raptor_ML} based location techniques are also our future plan.


\end{document}